\documentclass[10pt,conference]{IEEEtran}
\usepackage{cite}
\ifCLASSINFOpdf
   \usepackage[pdftex]{graphicx}
\else
\fi
\usepackage{amsmath}
\usepackage{algorithmic}
\usepackage{array,multirow}

\begin{document}
\title{The Intermittent Traveling Salesman Problem with Different Temperature Profiles: Greedy or not?}

\author{\IEEEauthorblockN{Pieter Leyman,
San Tu Pham and
Patrick De Causmaecker}
\IEEEauthorblockA{CODeS, Department of Computer Science \& imec-ITEC, KU Leuven KULAK\\ 
Etienne Sabbelaan 53, 8500 Kortrijk (Belgium)\\
Email: pieter.leyman@kuleuven.be, san.pham@kuleuven.be, patrick.decausmaecker@kuleuven.be}}

\maketitle

\begin{abstract}
In this research, we discuss the intermittent traveling salesman problem (ITSP), which extends the traditional traveling salesman problem (TSP) by imposing temperature restrictions on each node. These additional constraints limit the maximum allowable visit time per node, and result in multiple visits for each node which cannot be serviced in a single visit. We discuss three different temperature increase and decrease functions, namely a linear, a quadratic and an exponential function. To solve the problem, we consider three different solution representations as part of a metaheuristic approach. We argue that in case of similar temperature increase and decrease profiles, it is always beneficial to apply a greedy approach, i.e. to process as much as possible given the current node temperature.
\end{abstract}


\IEEEpeerreviewmaketitle

\section{\label{intro}Introduction}
In industry, drilling operations on pieces of material often increase the temperature of the component. In order to avoid overheating the material at any one point, the processing operation is split into multiple parts. This implies that nodes may need to be visited more than once. Additionally, there may be a waiting time at some nodes to be able to continue processing. Similar to \cite{pham2017}, we translate the problem to a traveling salesman problem (TSP) variant, with multiple visits per node. The resulting problem is called the intermittent traveling salesman problem (ITSP), due to the required time between different visits of a node to avoid overheating.

The TSP has been studied in depth in the field of combinatorial optimization \cite{johnson1997,gutin2007}, but several variants exist. To clearly distinguish the ITSP from existing extensions, we briefly discuss the most similar problems.
\begin{itemize}
\item The TSP with multiple visits (TSPM) \cite{gutin2007}: similar to the ITSP each node has to be visited several times, but unlike in the ITSP no time constraints exist between multiple visits. 
\item The TSP with time windows (TSPTW) \cite{dumas1995}: time windows within which the node has to be visited are associated to each node, but multiple visits are not required.
\item The inventory routing problem (IRP) \cite{campbell1998}: in the IRP, the goal is to find the best shipping policy to supply several retailers (nodes) with a common product subject to vehicle capacities, minimum required demand and limited storage capacity at the retailers. The major difference with the ITSP lies in the effect of traveling time on inventory levels. Whereas in the ITSP the temperature at a node decreases during travel time, in the IRP traveling routes compose one discrete period.
\end{itemize}

In the remainder of this paper, we first discuss the problem definition with different temperature profiles in section \ref{prob}, before going into detail about the proposed metaheuristic approach in section \ref{meth}. Preliminary results for the temperature profiles and metaheuristic variations are the focus of section \ref{res}, whereas we finish with a conclusion and future work in section \ref{concl}.

\section{\label{prob}Problem definition}
A network of nodes can be represented by an undirected graph $G=(N,A)$, with $N$ the nodes or points and $A$ the arcs or routes between different points. Each node $i$ ($i\in N=\{1,\ldots,n\}$) has a processing time $p_i$, and the distance between each pair of nodes $(i,j)$ is $d_{ij}$. It is explicitly assumed that $d_{ij}=d_{ji}$ and that the triangle inequality holds. Due to the temperature constraint (maximum temperature of $B$ at each node), nodes may have to be visited more than once. The goal is to minimize the total completion time, i.e. the time at which all nodes have been fully processed and the ``salesman'' has returned to its starting node. The objective function value consists of the total processing time of each node, the distances of the selected routes between the nodes, and any waiting time required in case the temperature at a node is too high but we want to process anyway (section \ref{meth}). As a result, the corresponding TSP without temperature constraints serves as a lower bound for the ITSP.

To model the temperature of a node $i$ at a time $t$, we use equations (\ref{cons}) and (\ref{temp}). The first equation determines the number of consecutive time units that have been processed for node $i$ at time $t$ ($c_{it}$), based on $y_{it}$ which is a binary variable equal to one if node $i$ is processed at time $t$ and zero otherwise. If job $i$ is processed at time $t$ equation (\ref{cons}) increases by one, whereas otherwise it decreases by one with a minimum value of zero for $c_{it}$. $c_{i0}$ is equal to zero for all nodes $i$.

\begin{multline}
	\label{cons}
	c_{it}=(c_{it-1}+1)\cdot y_{it}+max(c_{it-1}-1;0)\cdot (1-y_{it}), \\ 
	\forall{t}>0, \forall{i} \in N
\end{multline}

Equation (\ref{temp}) sets the temperature $B_{it}$ of a node $i$ at time $t$ based on the corresponding value of $c_{it}$. An increase function $f_1$ and a decrease function $f_2$, which determine the rate at which the temperature changes, are also defined. We employ three variants for the temperature profile changes, namely a linear ($f(t)=t$), a quadratic ($f(t)=t^2$) and an exponential ($f(t)=e^t$) function. In the linear variant the temperature increase or decrease is the same as the change in $c_{it}$, namely it increases or decreases by 1. For the quadratic case the temperature is the cube of the number of consecutively processed time units, and in case of the exponential function the base number $e$ is used with the time as exponent.

\begin{multline}
	\label{temp}
	B_{it}=f_1(c_{it})\cdot y_{it}+f_2(c_{it})\cdot (1-y_{it}), \forall{t}\ge0, \forall{i} \in N
\end{multline}

\begin{figure}
	\centering
	\includegraphics[width=9cm]{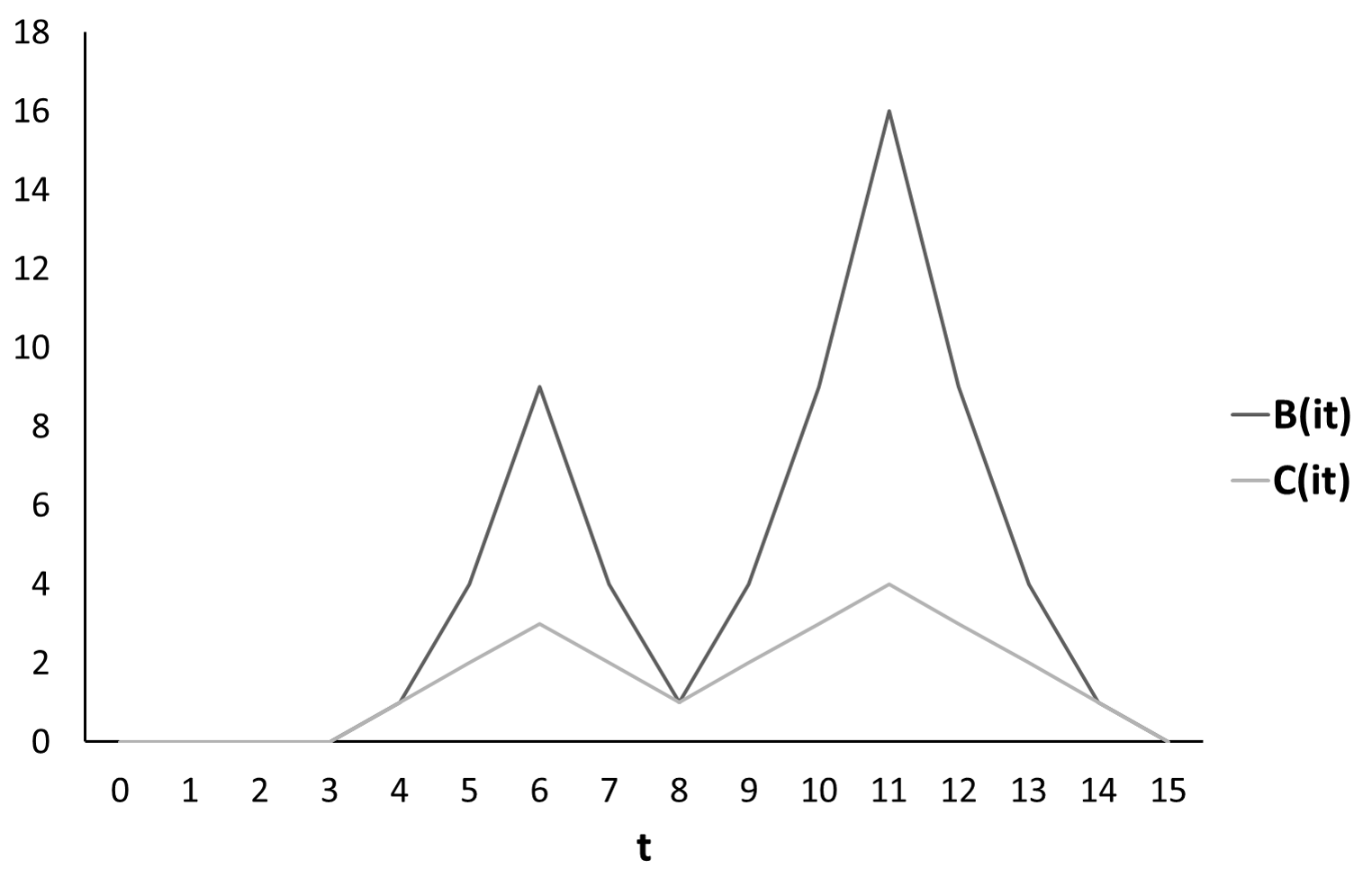}
	\caption{Example of profiles.}
	\label{fig1}
\end{figure}

Allow us to illustrate the application of both functions with a simple example. Assume that we have a node with a total processing time of 6, a quadratic increase and decrease temperature function, and that the maximum temperature $B$ equals 16. We start processing the job at time 3 and process for 3 time units until time 6 (node temperature equal to 9 or $3^2$), after which no processing occurs for 2 time units. We then process the remaining 3 time units until time 11 (node temperature of 16 or $(3-2+3)^2$). The resulting consecutive time units and temperature profiles are shown in Figure \ref{fig1}. We can observe that the temperature is indeed a function of the number of consecutively processed time units, and that the temperature decreases when no processing occurs.

Based on the three possible temperature increase and decrease profiles, nine combinations exist. In the remainder of this manuscript, we assume that the same function is used for both the increase and decrease of the temperature, but in our presentation we will discuss combinations as well.

\section{\label{meth}Methodology}
In this section, we discuss our solution approach for the ITSP. We propose to use three different applications of a genetic algorithm (GA) based on three different solution representations. We first go into detail about the representations, before giving an overview of the employed metaheuristic.

\subsection{\label{repr}Solution representation}
The choice of solution representation is important since we not only have to determine the order in which the different nodes are visited, but also the amount of time processed for each visit. Since, to the best of our knowledge, no previous research exists on the impact of solution representations for the ITSP, we employ three different approaches in order to determine their impact on the solution quality.

\begin{enumerate}
\item A single list (1L) representation consisting of a node list (NL). The NL contains the node numbers in the order in which they are processed, but due to the nature of the ITSP each node $i$ occurs $\left\lceil\frac{p_i}{\lfloor f_1^{-1}(B)\rfloor}\right\rceil$ times. This function determines the maximum required number of splits for each node, depending on $p_i$, $B$ and the temperature increase function $f_1(t)$. The maximum required number of splits corresponds with a greedy approach since in this case we choose to always process as much as possible during a visit given the current node temperature. Only for the final visit is waiting included if required. 
\item A two list (2L) representation with a NL and a processing time list (PTL). The NL again contains the order in which the nodes are processed, but each node $i$ occurs $p_i$ times instead of just a single time. The PTL then holds the actual processing times at each occurrence of a node $i$, with a total value over all occurrences equal to $p_i$. PTL values may be equal to zero to signify that less than $p_i$ splits are used for node $i$. As a result, these visits without processing time may increase the total duration because of distances traveled. 
\item A three list (3L) representation with a NL, a PTL and a split list (SL). The third list is the number of visits or splits for each node, and its value lies within the interval $\{1;p_i\}$ for each node $i$. A value of 1 means that this node is only visited once for a duration of $p_i$, whereas a value of $p_i$ implies as many visits each with a duration of 1. The NL and PTL are similar as for the second representation, with the major difference that both lists are only as long as the sum of the job splits in the SL. As a result, the PTL does not contain any zero values. Recall that in the second representation the PTL could have zero values and that both the NL and PTL sizes equal the sum of the job durations. It can be stated that the 3L representation constitutes a middle ground between the other two. It is less naive than the second one, since no unused visits with a processing time of zero are included. It does, however, allow for more and different splits than the greedy approach, by incorporating more information. 
\end{enumerate}

In the remainder of this paper we refer to 1L, 2L and 3L as three types of \textit{solution representations}, whereas NL, PTL and SL are \textit{lists} which can be part of the representations.

\begin{figure}
	\centering
	\includegraphics[width=5cm]{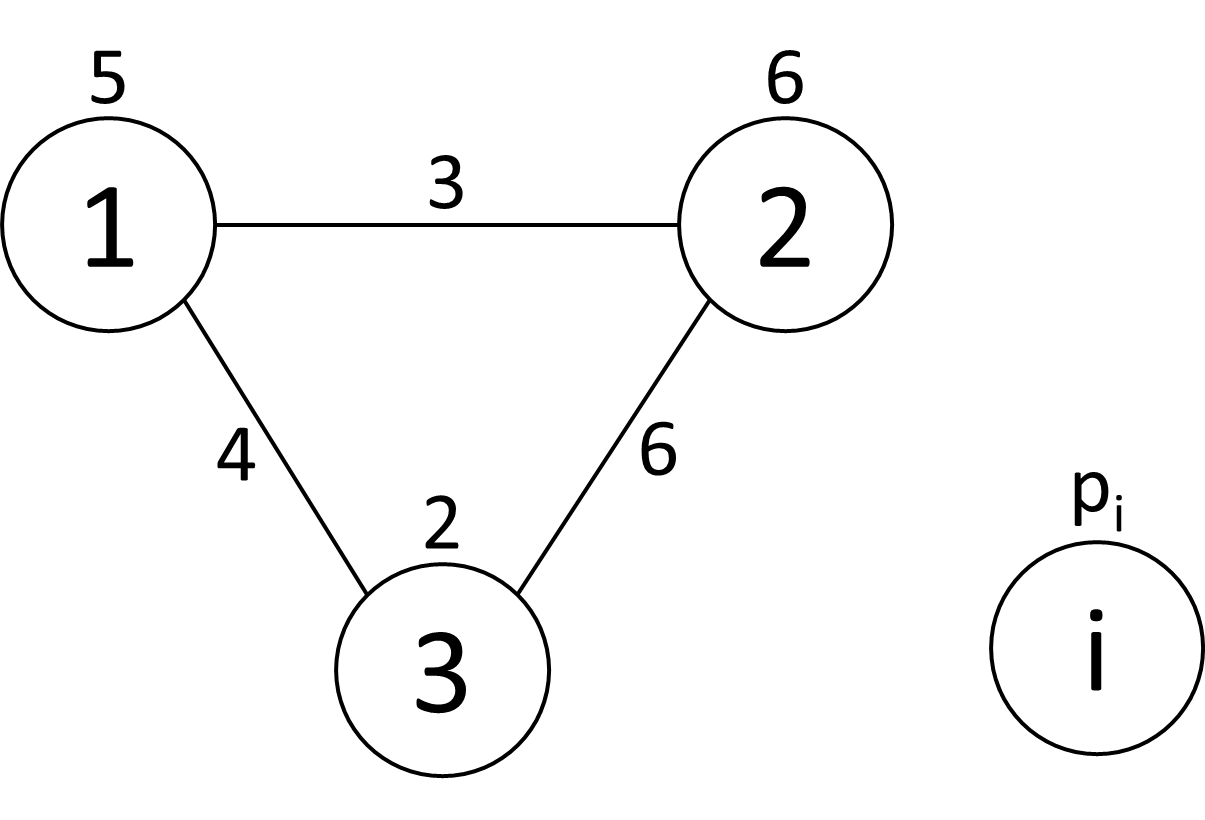}
	\caption{Example network.}
	\label{fig2}
\end{figure}

To illustrate the three representations, consider the example network in Figure \ref{fig2}, with three jobs and a maximum temperature $B$ of 3. We furthermore assume a linear temperature increase and decrease function for each job, which implies that for each job 3 time units can be processed consecutively without exceeding the maximum temperature.
\begin{enumerate}
\item With the single list we first calculate the number of splits based on the processing times and the maximum temperature. This results in 2 splits for job 1, 2 for job 2 and 1 for job 3. Assume that we have NL equal to (2, 1, 3, 1, 2). We process job 2 for a duration of 3 (the maximum amount possible), then process job 1 for 3 time units as well, move on to job 3 for a duration of 2, return to 1 for another 2 time units and finish with 3 time units for job 2. The total duration is equal to the sum of the job durations, the distances traveled and any waiting time. The first equals 13, the second 14 and the third 0 since no waiting times are needed. As a result, the total duration is equal to 27.
\item For the two list representation, assume a NL of (1, 1, 2, 2, 2, 1, 3, 3, 2, 2, 1, 1, 2) and a PTL of (2, 0, 1, 0, 1, 2, 2, 0, 1, 3, 0, 1, 0). Consider that the length of both lists is equal to 13, or to the sum of the individual job durations, that several zeros are included in the PTL, and that the sum of the corresponding visit durations equals $p_i$ for each job. The total duration is equal to 13 (the total job duration, the same as before) + 25 (the total distance traveled, including returning to node 1 where the tour started) + 0 (the total waiting time) = 38. Due to the increase in total distance the objective function for this NL and PTL combination is worse than for the greedy NL.
\item In case of the three lists, we use the NL (1, 2, 1, 2, 3, 2), PTL (3, 1, 2, 2, 2, 3) and SL (2, 3, 1). Each job is included as many times in the NL and PTL as the number of splits in the SL. The sum of the PTL values is again equal to $p_i$ for each of the three jobs. The total completion time equals 13 + 16 + 0 = 29.
\end{enumerate}
In the example we have explicitly assumed linear temperature profiles, which in this case resulted in no waiting time for the given lists. However, if the profiles would increase and decrease in a quadratic manner, then only 1 time unit can be processed before a node has to cool down. In the 3L example this results in repeatedly processing 1 time unit and waiting for 1 time unit. For instance, the first visit to node 1 involves 3 time units of processing time, which results in a total time of 5 (= 3 x 1 processing, 2 x 1 waiting). The same logic applies for the other visits in the tour. Hence, in the example the total duration increases from 29 to 36 due to 7 time units of waiting.

\subsection{\label{GA}Genetic algorithm}
An overview of our GA can be found in Figure \ref{figGA}. The population $P$ is initialized by generating $|P|$ random 1L, 2L or 3L respectively. This includes any relevant repair method and the evaluation of the solution value (cf. infra). The selection operator is the elite selection of \cite{leyman2015}, which selects one parent based on a four-tournament selection and the other one at random from the subset $R$ of best solutions in $P$. In the population update the best $|R|$ elements of the previous generation are retained, and the rest is replaced by the best children. Afterwards, the set $R$ is updated based on the new $P$. It has been shown that due to the elite selection and population update, the GA contains elements of both scatter search and evolutionary path relinking \cite{leyman2017}. However, for the sake of simplicity we refer to the algorithm as a GA, although the more general term evolutionary algorithm (EA) would also be correct.

\begin{figure}[!t]
\begin{algorithmic}[1]
\STATE \textit{Initialize} population $P$
\vspace{-0.5mm}
\STATE \textbf{Repeat}
\vspace{-0.5mm}
\STATE \hspace{\algorithmicindent}\textit{Select} $p_1$ and $p_2$ from $P$
\vspace{-0.5mm}
\STATE \hspace{\algorithmicindent}\textit{Crossover*} $p_1$ and $p_2$ $\rightarrow$ $o_1$ and $o_2$
\vspace{-0.5mm}
\STATE \hspace{\algorithmicindent}\textit{Repair*} $o_1$ and $o_2$ 
\vspace{-0.5mm}
\STATE \hspace{\algorithmicindent}\textbf{For} each offspring $o$ \textbf{do}
\vspace{-0.5mm}
\STATE \hspace{\algorithmicindent}\hspace{\algorithmicindent}\textit{Mutate*} $o$ 
\vspace{-0.5mm}
\STATE \hspace{\algorithmicindent}\hspace{\algorithmicindent}\textit{Repair*} $o$ 
\vspace{-0.5mm}
\STATE \hspace{\algorithmicindent}\hspace{\algorithmicindent}\textit{Evaluate*} $o$ 
\vspace{-0.5mm}
\STATE \hspace{\algorithmicindent}\textbf{End for}
\vspace{-0.5mm}
\STATE \hspace{\algorithmicindent}\textit{Update} $P$: remove $|P|-|R|$ worst, add $|R|$ best
\vspace{-0.5mm}
\STATE \hspace{\algorithmicindent}offspring, update $R$
\vspace{-0.5mm}
\STATE \textbf{Until} stopping criterion met
\caption{GA outline}\label{figGA}
\end{algorithmic}
\end{figure}

The asterisks (*) indicate that the corresponding steps differ based on the solution representation used (1L, 2L or 3L), since these parts require different applications of some operators. We also distinguish between the three types of lists used (NL, PTL and SL) when discussing the operators with an * in more detail, since it are these lists which result in differences between the operators used. Recall that 1L consists only of a NL, 2L of a NL and PTL, and 3L of a NL, PTL and SL. In Table \ref{tabop} an overview of the differences in terms of operators between the NL, PTL and SL is displayed, whereas Table \ref{tabop2} shows the repair and solution evaluation employed for the three solution representations. 
\begin{itemize}
\item NL: We apply a one-point crossover and a two-activity swap as crossover and mutation operator with a mutation rate of $m_{NL}$ respectively. No repair method is required, and the evaluation consists of processing as much as possible of a node during each visit. Only the final visit to a node may include waiting time to ensure that the temperature constraint is not violated.
\item PTL: We again use a one-point crossover. Consider that the NL and PTL always have the same length, so the same crossover point can be used for both lists. However, the PTL requires a repair method to ensure that for each job the sum of the corresponding PTL values equals $p_i$. In case the sum is too large (small), this repair method randomly decreases (increases) PTL values of the job $i$ until the condition is met. In terms of mutation, we have chosen a random change operator which randomly selects a different PTL value. The mutation rate $m_{PTL}$ is used for each job in the list, rather than for the PTL as a whole. Also after the mutation, we have to apply the same repair method to ensure that the PTL is feasible. Finally, in terms of solution evaluation we process the corresponding values in the PTL during each visit to a node. Hence, waiting time is included if this is required based on the PTL value. 
\item SL: The crossover operator is an adjusted version of the one-point crossover due to the different length of the SL on the one hand and the NL and PTL on the other hand, and uses a different crossover point. Additionally, the crossover of NL and PTL takes the different lengths of the parents' lists into account. The mutation of the SL is similar to that of the PTL; a random change is applied to each job with a rate of $m_{SL}$. The use of a SL requires a second repair method to ensure that the number of splits of a job in the SL corresponds with the number of job occurrences in the other two lists. This method adjusts the NL and PTL by removing (adding) job occurrences in the NL and increasing (decreasing) values in the PTL at random. This way the total length of both lists corresponds with the sum of the SL values. The repair method is applied after both the crossover and the mutation. Finally, the usage of a SL (3L representation) has no effect on the evaluation of the solution, so the same PTL-based technique can be used as without a SL (2L representation).
\end{itemize}

\begin{table}[!t]
\renewcommand{\arraystretch}{1.3}
\caption{GA Operators.}
\label{tabop}
\small
\begin{center}
\begin{tabular}{|c|c|c|c|}
\hline
\textbf{Operator} & \textbf{NL} & \textbf{PTL} & \textbf{SL}\\
\hline
Crossover & 1-point & 1-point & 1-point'\\
Mutation & 2-swap & Random change & Random change\\
\hline
\end{tabular}
\end{center}
\end{table}

\begin{table}[!t]
\renewcommand{\arraystretch}{1.3}
\caption{Repair \& Solution Evaluation.}
\label{tabop2}
\small
\begin{center}
\begin{tabular}{|c|c|c|c|}
\hline
 & \textbf{1L} & \textbf{2L} & \textbf{3L}\\
\hline
Lists & NL & NL, PTL & NL, PTL, SL\\
Repair & / & Repair 1 & Repair 1 \& 2\\
Evaluation & Greedy & PTL-based & PTL-based\\
\hline
\end{tabular}
\end{center}
\end{table}

\section{\label{res}Results}
To test our approach, we have generated our own data. Table \ref{tabparam} shows the selected values for different data parameters. For each combination we generate 10 variations, which results in a total of 400 (= 10 x 2 x 2 x 2 x 5) instances. In terms of temperature profiles, we test the three variants discussed, and assume that the increase and decrease functions of each node are the same, i.e. linear, quadratic or exponential. We use a stopping criterion of 5000 tours for each instance. 

\begin{table}[!t]
\renewcommand{\arraystretch}{1.3}
\caption{Data Parameters.}
\label{tabparam}
\small
\begin{center}
\begin{tabular}{|c|c|}
\hline
\textbf{Parameter} & \textbf{Values}\\
\hline
Number of nodes ($|N|$) & 10, 50\\
Processing time ($p_i$) & [10;20], [10;100]\\
Distance ($d_{ij}$) & [10;20], [10;100]\\
Maximum temperature ($B$) & 20, 40, 60, 80, 100\\
\hline
\end{tabular}
\end{center}
\end{table}

In Table \ref{tabparamalg} we show the best found parameter values for each of the metaheuristic's parameters. We distinguish between the three solution representations 1L, 2L and 3L. We can conclude that the values for the population size $P$ and retention rate $R$ are in line with those of \cite{leyman2015}, whereas the mutation rates depend on the solution representation used. As more lists are included in a representation, our results indicate that it is better to use lower mutation rates. 

\begin{table}[!t]
\renewcommand{\arraystretch}{1.3}
\caption{Algorithm Parameters.}
\label{tabparamalg}
\small
\begin{center}
\begin{tabular}{|c|c|c|c|}
\hline
\textbf{Parameter} & \textbf{1L} & \textbf{2L} & \textbf{3L}\\
\hline
$P$ & 50 & 50 & 50\\
$R$ & 5 & 5 & 5\\
$m_{NL}$ & 0.90 & 0.90 & 0.10\\
$m_{PTL}$ & / & 0.05 & 0.02\\
$m_{SL}$ & / & / & 0.01\\
\hline
\end{tabular}
\end{center}
\end{table}

The comparative results of the three different solution representations are displayed in Table \ref{tabalgcomp}, in terms of the average total duration (\textit{AvDur}) required to complete the processing of the material. We distinguish between the three solution representations and between the three temperature profiles linear (L), quadratic (Q) and exponential (E). Two main conclusions can be drawn based on the table. 

First, there is a large difference in performance between the three representations. 1L always has the best performance, whereas 2L always performs worst. The explanation for these results lies in the size of the solution space employed in the GA. Especially in the 2L representation, a huge number of possible (feasible) combinations exist. For instance in case of 10 jobs with each a duration of 10, both the NL and PTL already have a length of 100, whereas the 1L may be as short as 10. This comparative result confirms that it is indeed best to apply a greedy approach in case of similar temperature increase and decrease profiles. However, given the results for 3L, it can be stated that the latter representation should be investigated further, since it shows potential. In particular in case of different increase and decrease profiles, the 3L may prove a valuable alternative to the 1L.

Second, the increases in temperature profiles have a profoundly negative impact on the total duration. This observation is important from a managerial perspective, since it can be used to show that operations with a limited temperature increase and decrease are preferable compared to those with a more steeper increase and decrease.

\begin{table}[!t]
\renewcommand{\arraystretch}{1.3}
\caption{Comparison of solution representations (\textit{AvDur}).}
\label{tabalgcomp}
\small
\begin{center}
\begin{tabular}{|c|c|c|c|}
\hline
\textbf{Function} & \textbf{1L} & \textbf{2L} & \textbf{3L}\\
\hline
L & \textbf{1089.62} & 10,866.05 & 1743.50\\
Q & \textbf{2076.63} & 13,564.22 & 3037.27\\
E & \textbf{3521.63} & 16,620.25 & 4672.14\\
\hline
\end{tabular}
\end{center}
\end{table}

\section{\label{concl}Conclusions \& future work}
In this paper, we have discussed the intermittent traveling salesman problem (ITSP). This problem imposes additional temperature constraints on the nodes in the network, such that the maximum allowable processing time on a node is limited. We have employed a linear, a quadratic and an exponential function for the temperature increase and decrease profiles. A metaheuristic which employs three different solution representations has been proposed, and it was shown that in case the increase and decrease functions for the node temperatures are the same, it is best to apply a greedy approach and process as much as possible.

In the future, we aim to investigate the impact if different temperature increase and decrease profiles are used. We expect that as long as the temperature in a node always decreases at least as fast as it increases, it will be beneficial to process as much as possible during each visit. However, in case the temperature decreases slower than it increases it becomes less obvious what is best course of action. Table \ref{tabfuture} shows the (expected) impact of temperature profiles for linear (L), quadratic (Q) and exponential (E) functions.

\begin{table}[!t]
\renewcommand{\arraystretch}{1.3}
\caption{Impact temperature profiles.}
\label{tabfuture}
\small
\begin{center}
\begin{tabular}{|cc|c|c|c|}
\hline
& & \multicolumn{3}{c|}{\textbf{Decrease}}\\\cline{3-5}
& & L & Q & E\\
\hline
& L & \textbf{Greedy} & Greedy & Greedy\\
\textbf{Increase} & Q & ? & \textbf{Greedy} & Greedy\\
& E & ? & ? & \textbf{Greedy}\\
\hline
\end{tabular}
\end{center}
\end{table}

A second interesting research avenue concerns the shape of the temperature functions. In this manuscript we explicitly assumed that these functions were straightforward (i.e. $f(t)=t$, $f(t)=t^2$ and $f(t)=e^t$), which resulted in a clear distinction between them, since e.g. the linear function was never larger than the quadratic one for the same value of $t$. However, employing functions such as $f(t)=4\cdot t$ (increase) and $f(t)=t^2/2$ (decrease) would make it harder to determine how much should be best processed during a visit given the current node temperature.

Finally, we currently assume a uniform surface and density of the material (section \ref{intro}), where all the points or nodes are similar. It might prove interesting to investigate the impact of the shape and density of the material surface, and what the link might be with the temperature profiles.

\vspace{1.0mm}
\textbf{Acknowledgements:} This research was supported by the Belgian Science Policy Office (BELSPO) in the Interuniversity Attraction Pole COMEX.


\end{document}